\documentclass[aps,prl,twocolumn,superscriptaddress]{revtex4-2}
\usepackage{multirow}
\usepackage{graphicx}    
\usepackage{amsmath}     
\usepackage{amssymb}     
\usepackage{hyperref}    
\usepackage{times}       
\usepackage{physics}     
\usepackage{color}

\begin{document}

\preprint{APS/123-QED}
\title{Polarization–Magnetization Coupling in Visible-Light Ferroelectric Double Perovskites}
\author{Sathiyamoorthy Buvaneswaran}
\affiliation{Department of Physics and Nanotechnology,
Faculty of Engineering and Technology, SRM Institute of Science and Technology, Kattankulathur 603 203, Tamil Nadu, India}
\author{Trilochan Sahoo}
\affiliation{Department of Physics and Nanotechnology,
Faculty of Engineering and Technology, SRM Institute of Science and Technology, Kattankulathur 603 203, Tamil Nadu, India}
\author{Saurabh Ghosh}
\email{saurabhghosh2802@gmail.com}
\affiliation{Department of Physics and Nanotechnology,
Faculty of Engineering and Technology, SRM Institute of Science and Technology,
Kattankulathur 603 203, Tamil Nadu, India}
\affiliation{Center for Advanced Computational and Theoretical Sciences (SRM-ACTS), Faculty of Engineering and Technology,SRM Institute of Science and Technology, Kattankulathur 603 203, Tamil Nadu, India}

\date{\today}
\begin{abstract}
The bulk photovoltaic effect (BPVE), arising from broken inversion symmetry in ferroelectrics, offers a distinct pathway toward high-efficiency next-generation photovoltaics. We propose and investigate A/A$^\prime$-ordered double perovskites KLaFeMoO$_6$ and NaLaFeMoO$_6$ as promising single-phase ferroelectric photovoltaic (FE-PV) materials. First-principles calculations reveal robust P2$_1$ symmetry with A-site layer and B-site rock-salt ordering, accompanied by hybrid improper ferroelectricity driven by $a^{-}a^{-}c^{+}$ octahedral tilts. Both compounds exhibit significant spontaneous polarization and indirect band gaps of $\sim$ 1.8 eV, well suited for visible-light absorption ($>$10$^5$ cm$^{-1}$). Low carrier effective masses along the polar axis indicate efficient charge transport. \textit{Ab initio} molecular dynamics simulations (AIMD) show that polarization-coupled magnetization switching is feasible above room temperature, making these materials suitable for room-temperature applications.
\end{abstract}
\maketitle
The BPVE, which arises in noncentrosymmetric single-phase materials, enables the generation of photocurrent without requiring a p–n junction \cite{kraut1979anomalous,von1981theory,bhatnagar2013role}. In such polar systems, the intrinsic electric field associated with spontaneous polarization facilitates the separation of photoexcited electron–hole pairs. Unlike conventional photovoltaic mechanisms, BPVE can yield open-circuit voltages exceeding the electronic band gap \cite{yang2010above,seidel2011efficient}, offering a potential route to surpass the Shockley–Queisser limit \cite{spanier2016power}. Ferroelectric perovskites such as BaTiO$_3$, LiNbO$_3$, and Pb(Zr,Ti)O$_3$, etc., have been extensively studied in this context \cite{ichiki2004photovoltaic,fridkin1978anomalous,glass1995high,dubovik1995bulk}; however, their wide band gaps $>$ 3 eV limit absorption to the ultraviolet region, thus constraining their photovoltaic performance. Among them, BiFeO$_3$ has attracted particular attention due to its large spontaneous polarization and high Curie temperature (T$_c$), but its relatively large bandgap ($\sim$ 2.67 eV) and weak absorption in the visible range lead to limited photocurrent generation in practical applications \cite{choi2009switchable,basu2008photoconductivity,ghosh2019electronic}. These limitations highlight the need for novel ferroelectric materials that combine large polarization, an appropriate bandgap, strong visible-light absorption, efficient charge-transport characteristics, and robust ferroelectricity sustained at or above room temperature. 
\par
Double perovskite oxides (DPOs), with the chemical formula AA$^\prime$BB$^\prime$O$_6$, are promising candidates in this context. Their structural and functional properties in DPOs are governed by collective lattice distortions and symmetry-breaking couplings among key modes—namely, in-phase ($Q_{R+}$, $a^{0}$$a^{0}$$c^{+}$) and out-of-phase ($Q_{R-}$, $a^{0}$$a^{0}$$c^{-}$) octahedral rotations, in-plane tilts ($Q_T$, $a^{-}$$a^{-}$$c^{0}$), and polar A-site displacements ($Q_{\mathrm{AFEA}}$) \cite{benedek2011hybrid,gayathri2023switching,cmpredictive}. The most common cation configuration features rock-salt ([R]) ordering on the B/B$^\prime$ sublattice and layered ([L]) or rock-salt ordering on the A/A$^\prime$ sublattice. When A-site [L] and B-site [R] orderings coexist with an $a^{-}$$a^{-}$$c^{+}$ tilt pattern, the lattice symmetry is reduced to a polar $P2_1$ phase. This symmetry breaking is driven by a trilinear coupling between the nonpolar $Q_T$ and $Q_{R+}$ modes and the polar $Q_{\mathrm{AFEA}}$ mode—characteristic of hybrid improper ferroelectricity (HIF). Thus, identifying DPOs with stable AA$^\prime$-site ordering, strong spontaneous polarization, and reduced bandgap opens up avenues to design novel visible-light-absorbing FE-PV materials.
\par
In this work, we investigate KLaFeMoO$_6$ and NaLaFeMoO$_6$ DPOs as promising FE-PV materials capable of operating at or above room temperature. DFT calculations reveal a polar $P2_1$ ground state with A-site [L] and B-site [R] ordering and an $a^{-}a^{-}c^{+}$ octahedral tilt pattern. Phonon symmetry-mode analysis identifies three key distortion modes ($Q_{R+}$, $Q_T$, $Q_{AFEA}$) that drive inversion symmetry breaking via the HIF mechanism, yielding robust polarization along the $b$-axis. Electronic structure calculations reveal indirect band gaps in the optimal $\sim$1.8 eV range, strong visible-light absorption ($>$10$^5$ cm$^{-1}$), and low carrier effective masses along the polar $b$ axis, favorable for carrier mobility. AIMD simulations show that the polarization coupled magnetization switching exceeds room temperature, confirming the thermal stability and suitability of the compounds for room-temperature multiferroic and optoelectronic applications. We have discussed how behaviour of the real part of the dielectric function can detect polarization-coupled magnetization switching. 
\par
\textit{\textbf{Computational details}}: First-principles density functional theory (DFT) calculations \cite{hohenberg1964density,kohn1965self} were performed using the full-potential linearized augmented plane wave (FP-LAPW) method in the WIEN2k package \cite{blaha1990full}. Structural relaxations employed the GGA-PBE functional \cite{perdew1996generalized} with a $\Gamma$-centered $6 \times 6 \times 4$ $k$-point mesh. On-site Coulomb interactions for transition metal $d$ orbitals were treated via the GGA+$U$ scheme with Hubbard parameter $U_{\mathrm{eff}} = U - J_H$ of 4.5 eV for Fe ($3d$) and 1.0 eV for Mo ($4d$)\cite{dudarev1998electron}. Electronic structure and band gaps were further refined using the Modified Becke-Johnson (MBJ) potential \cite{tran2009accurate} with spin-orbit coupling (SOC) along all three crystallographic directions, considering the low-spin configurations along the $b$-axis. Geometry optimizations were converged when the total energy change was below $10^{-8}$ eV and Hellmann–Feynman forces were below 0.001 eV Å$^{-1}$. Electronic and optical properties were computed using GGA+U+MBJ+SOC on a $12 \times 12 \times 8$ $k$-point mesh. The reliability of this approach for similar systems has been validated in our previous studies \cite{buvaneswaran2023design,buvaneswaran2025ferroelectric}.
\begin{figure}
\centering
\includegraphics[width=\linewidth]{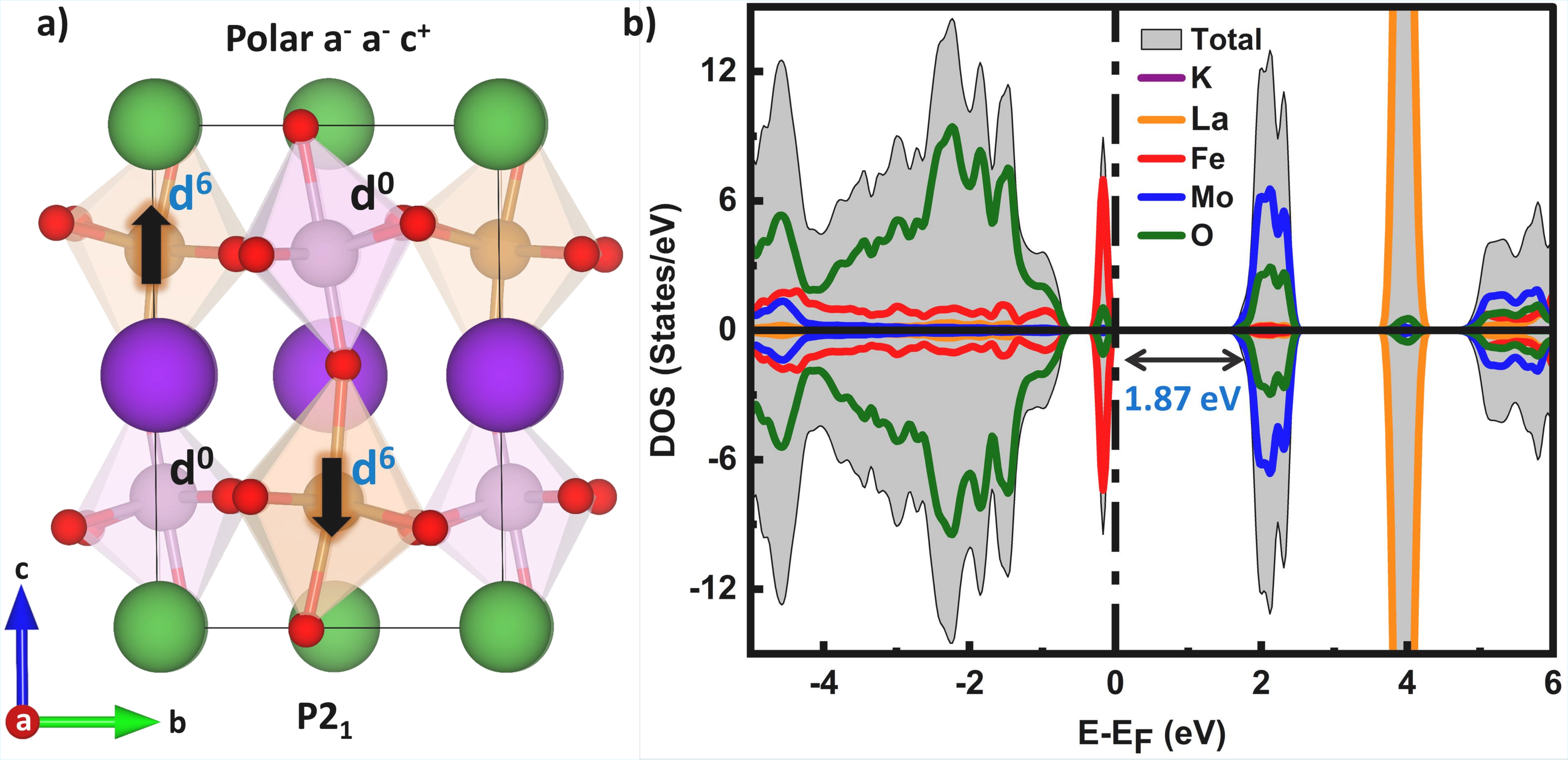}\vspace{-0pt}
\caption {Optimized $P2_1$ structure of KLaFeMoO$_6$ showing A-type AFM ordering, in which Fe spins are ferromagnetically aligned within the $ab$ plane and antiferromagnetically coupled along the $c$ axis. (b) Corresponding total DOS of KLaFeMoO$_6$.} 
\label{Figure1}
\end{figure} 
\par
\textit{\textbf{Origin of ferroelectricity}}: The compound Sr$_2$FeMoO$_6$ (SFMO) in the tetragonal $I4/mmm$ phase is a well-known half-metallic ferromagnet, exhibiting robust Fe/Mo rock-salt ordering and a high T$_c$ ($\approx$ 410–450 K) \cite{philipp2003structural,tomioka2000magnetic}. In this phase, Fe$^{3+}$ (S = 5/2) and Mo$^{5+}$ (S = 1/2) sublattices are ferromagnetically aligned within their own sublattice, while the two sublattices are antiferromagnetically coupled to each other \cite{kobayashi1998room,sarma2001new}. These features make SFMO a benchmark system in spintronics and an excellent electronic platform for designing multifunctional materials. The optimized structure with spin configuration and the calculated density of states (DOS) of SFMO, consistent with earlier reports \cite{kumar2012ferrimagnetism,sarma2001properties,ray2001electronic}, are shown in \textcolor{blue}{Fig. S1} of the Supplemental Material (SM) \cite{SM}. We select K$^{1+}$ and La$^{3+}$ at the (A, A$^\prime$)-site to replace Sr$^{2+}$ on this SFMO parent structure. Our previous studies indicated that the large charge and size contrast between A and A$^\prime$ promotes strong A-site cation ordering, which is essential for stabilizing polar distortions and enabling HIF in DPOs \cite{shaikh2021investigation}. 
\par
To induce ferroelectricity in DPOs, two key structural conditions must be satisfied: (i) stabilization of the $a^{-}a^{-}c^{+}$ tilt pattern within the polar $P2_{1}$ phase and (ii) formation of A/A$^\prime$-site layered ordering. The $a^{-}a^{-}c^{+}$ tilt configuration breaks inversion symmetry, while the A/A$^\prime$ cation ordering further couples with octahedral rotations to generate spontaneous polarization via the HIF mechanism. DFT total energy calculations confirm that the $P2_{1}$ phase with an $a^{-}a^{-}c^{+}$ tilt pattern is the most stable configuration among various possible tilt distortions. To evaluate cation ordering tendencies, special quasi-random structure (SQS) models \cite{zunger1990special} were employed, which reveal that the ordered arrangement is energetically more favorable than disordered configurations. Combined DFT and SQS results thus confirm that KLaFeMoO$_6$ stabilizes in a polar $P2_{1}$ phase with an $a^{-}a^{-}c^{+}$ tilt pattern and A-site [L] and B-site [R] ordering. A detailed explanation of the SQS calculation method and results is provided in SM \textcolor{blue}{Note S1} \cite{SM}. 
\par
To elucidate the origin of ferroelectricity in KLaFeMoO$_6$, we performed functional mode analysis by decomposing the low-symmetry polar $P2_1$ structure with respect to the high-symmetry cubic $P4/nmm$ reference phase (SM \textcolor{blue}{Fig. S3 (a,b)}). Phonon mode decomposition identifies four dominant lattice distortions: in-plane octahedral tilting ($Q_T$), out-of-plane rotation ($Q_{R+}$), two-dimensional charge disproportionation ($Q_{\mathrm{CD_{2D}}}$), and polar A-site displacement ($Q_{\mathrm{AFEA}}$), as illustrated in SM \textcolor{blue}{Fig. S3 (c–f)} \cite{SM}. The corresponding group–subgroup relationship further outlines the symmetry-lowering pathway through several intermediate phases, as depicted in SM \textcolor{blue}{Fig. S3 (g)} \cite{SM}. A trilinear coupling among the $Q_T$, $Q_{R+}$, and $Q_{\mathrm{AFEA}}$ modes stabilizes the ferroelectric $P2_1$ phase through a HIF mechanism, resulting in a polarization along the b axis ($\vec{P}_{y}$). The calculated $\vec{P}_{y}$ of KLaFeMoO$_6$ is $\sim$14.42 $\mu$C/cm$^2$ using the Berry phase method \cite{king1993theory}.
\par
In the ground state, Fe ions are in the \(2+\) oxidation state with a high-spin \(3d^6\) configuration \(\left(\text{Fe}^{2+}: t_{2g}^4 e_g^2,\, S = 2\right)\), while Mo ions adopt a \(6+\) oxidation state corresponding to an empty \(4d^0\) configuration \(\left(\text{Mo}^{6+}: t_{2g}^0 e_g^0,\, S = 0\right)\). To identify the magnetic ground state of KLaFeMoO$_6$ we compared the total energies of different magnetic orderings, including ferromagnetic (FM) and A-, C-, and G-type antiferromagnetic (AFM) configurations. The lowest-energy state corresponds to A-type AFM ordering, in which Fe moments align ferromagnetically within the $ab$-plane and couple antiferromagnetically along the $c$-axis as shown in \textcolor{blue}{Fig. \ref{Figure1}a}. In this configuration, Fe exhibits a strong spin moment (3.73~\(\mu_\mathrm{B}\)/Fe\(^{2+}\)) and Mo develops a weaker spin moment (0.03~\(\mu_\mathrm{B}\)/Mo\(^{6+}\)) due to exchange-induced polarization effect. 
\par
To further explore the effect of A-site substitution, we replaced K with Na in the KLaFeMoO$_6$ lattice. The resulting NaLaFeMoO$_6$ stabilizes in the same polar P2$_1$ (a$^{-}$a$^{-}$c$^{+}$). The calculated polarization increases to $\sim$21.32 $\mu$C/cm$^2$ (with full path polarization calculation), attributed to the smaller ionic radius mismatch between Na$^{1+}$ and La$^{3+}$. It also retains A-type AFM ordering, with Fe and Mo magnetic moments slightly modified due to changes in the Fe–O–Mo bond angle and bond length. The optimized structure and corresponding DOS are shown in \textcolor{blue}{Fig. S4}. The structural and magnetic parameters are compared in \textcolor{blue}{Table S1} \cite{SM}.
\begin{figure}
\centering
\includegraphics[width=\linewidth]{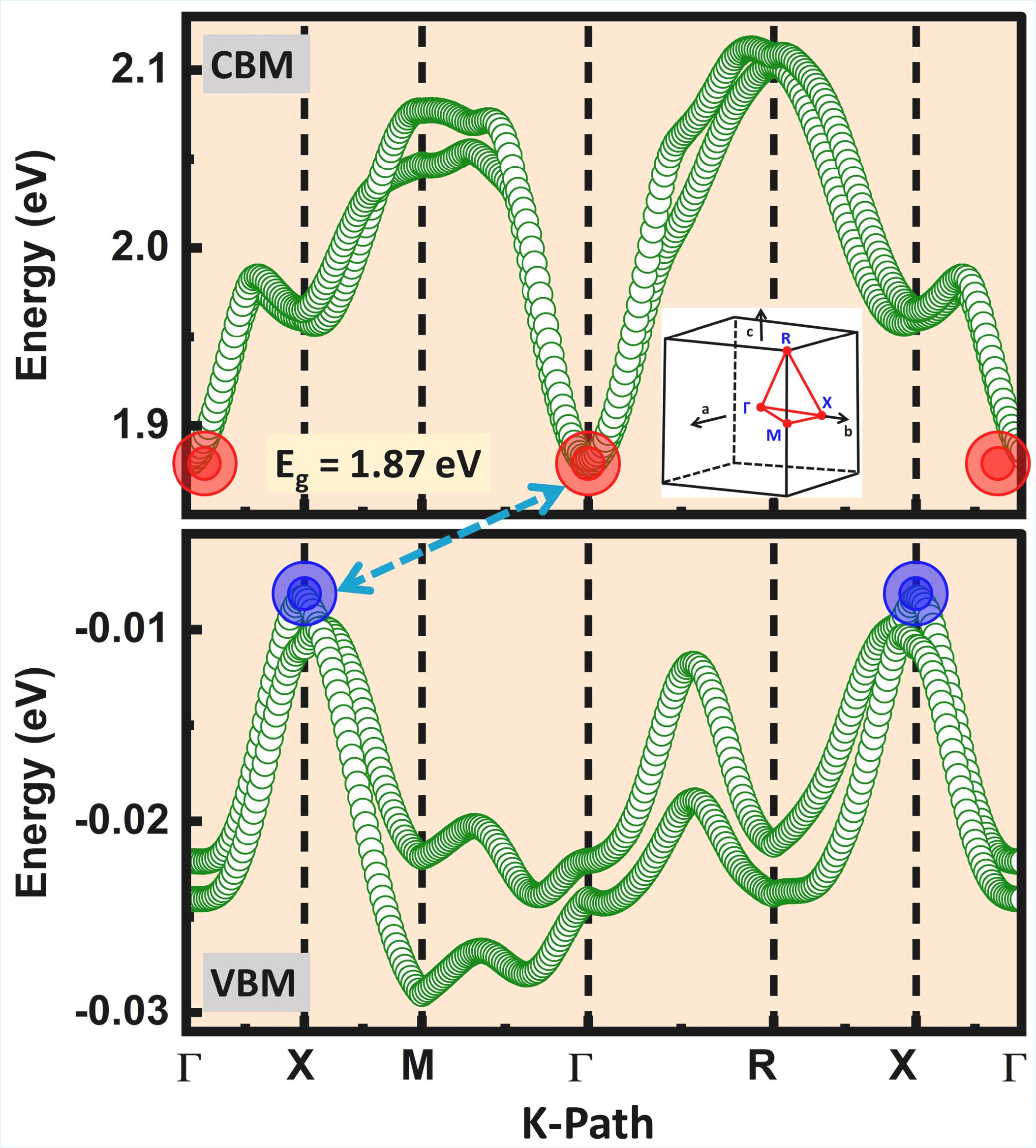}\vspace{-0pt}
\caption{Band dispersion of the highest VB and the lowest CB along high-symmetry directions in the Brillouin zone of KLaFeMoO$_6$, calculated using GGA+U+MBJ+SOC. Blue and red circles indicate the maxima of VB and minima of CB, respectively.}
\label{Figure2}
\end{figure} 
\par
\textit{\textbf{Electronic structure and band dispersion}}: From the above studies, we confirm that both materials exhibit strong ferroelectric polarization, which facilitates efficient separation of photoexcited carriers. The DOS profile of KLaFeMoO$_6$, as depicted in \textcolor{blue}{Fig. \ref{Figure1}b}, reveals that the valence band (VB) is primarily governed by strong Fe-$3d$ and O-$2p$ hybridization, while the conduction band (CB) mainly originates from overlapping Mo-$4d$ and O-$2p$ states. Such transition-metal–oxygen hybridization promotes delocalized electronic states, thereby enhancing carrier mobility and reducing recombination. The electronic structure indicates that KLaFeMoO$_6$ is an indirect band-gap semiconductor with an energy gap of 1.87 eV. Although an indirect band gap is less favorable for light absorption, it can reduce radiative electron–hole recombination rate, leading to longer carrier lifetimes ($\tau$) and greater diffusion lengths \cite{johnston2016hybrid}. The strong Fe–O and Mo–O covalency near the Fermi level in the VB and CB produces a high density of states, which can enhance optical transitions, increase the absorption coefficient in the visible region, and thereby boost photocurrent generation. Similarly, DOS profile of NaLaFeMoO$_6$, as shown in SM \textcolor{blue}{Fig. S4b} \cite{SM}, displays an indirect band gap of 1.82 eV, indicating that A-site chemistry has minimal influence on the energy gap since A-site states lie far from the Fermi level. The slight variation arises from subtle modifications in Fe–O–Mo bond geometry, consistent with earlier reports \cite{buvaneswaran2023design,buvaneswaran2025ferroelectric}.
\begin{table} 
\centering
\caption{Calculated hole ($m_h^*$) and electron ($m_e^*$) effective masses (in units of electron mass $m_e$) for AA$^\prime$FeMoO$_6$ (AA$^\prime$= KLa and NaLa) along different high-symmetry directions.}
\begin{tabular}{c |c |c|c| c| c}
\hline
\multirow{2}{*}{Direction} & \multicolumn{2}{c|}{\boldmath{$m_e^*$}} &
\multirow{2}{*}{Direction} & \multicolumn{2}{c}{\boldmath{$m_h^*$}} \\
 &  KLa & NaLa & & KLa & NaLa \\
\hline
X-$\Gamma$-X   & 0.55 & 0.63 & $\Gamma$-X-$\Gamma$ & 1.91 & 1.98 \\
M-$\Gamma$-M   & 0.59 & 0.68 & M-X-M & 2.53 & 2.61 \\
R-$\Gamma$-R   & 0.64 & 0.71 & R-X-R & 2.83 & 2.93 \\
\hline                  
\end{tabular}
\label{Table 1}
\end{table}
\par
The band dispersion of the highest VB and lowest CB of KLaFeMoO$_6$, calculated using GGA+U+MBJ+SOC, is shown in \textcolor{blue}{Fig. 2}. The band gap separates states formed by hybridized Fe–3$d$/O–2$p$ (VB) and Mo–4$d$/O–2$p$ (CB) orbitals, both exhibiting significant dispersion near the Fermi level, indicative of low carrier effective masses (m$^*$). Such dispersive features are beneficial for charge transport, as carrier mobility $\mu$ $\alpha$ 1/m$^*$. The band structure of NaLaFeMoO$_6$ is provided in SM \textcolor{blue}{Fig. S5} \cite{SM}. To quantify these features, we calculated the m$^*$ value of both NaLaFeMoO$_6$ and KLaFeMoO$_6$ along different high-symmetry directions in the Brillouin zone ( inserted in \textcolor{blue}{Fig. 2}), with the values summarized in \textcolor{blue}{Table \ref{Table 1}}.
\par
The electron (m$_e^*$) and hole (m$_h^*$) effective masses were obtained from the curvature of the CB minimum (Red circle) and VB maximum (Blue circle), respectively, as indicated in the band dispersion plot. These values are comparable to those of well-known BPVE materials, including BiFeO$_3$ (m$_e^*$ = 0.691 m$_e$, m$_h^*$ = 3.171 m$_e$), KNbO$_3$ (m$_e^*$ = 1.559 m$_e$, m$_h^*$ = 2.743 m$_e$), and tetragonal BaTiO$_3$ (m$_e^*$ = 0.482 m$_e$, m$_h^*$ = 1.082 m$_e$). This comparison highlights that both NaLaFeMoO$_6$ and KLaFeMoO$_6$ possess carrier transport characteristics on par with well established BPVE systems. Notably, the lowest effective masses were found along the polar b-axis (X–$\Gamma$–X), a direction of particular importance since previous experiments have shown that photo-induced output voltage V$_{OC}$ is observed only along the polarization axis (Perpendicular to the domain wall) \cite{yang2010above}. Thus, the combination of anisotropic band dispersion and low effective masses strongly suggests that these compounds are highly promising for efficient FE-PV applications.
\begin{figure}
\centering
\includegraphics[width=\linewidth]{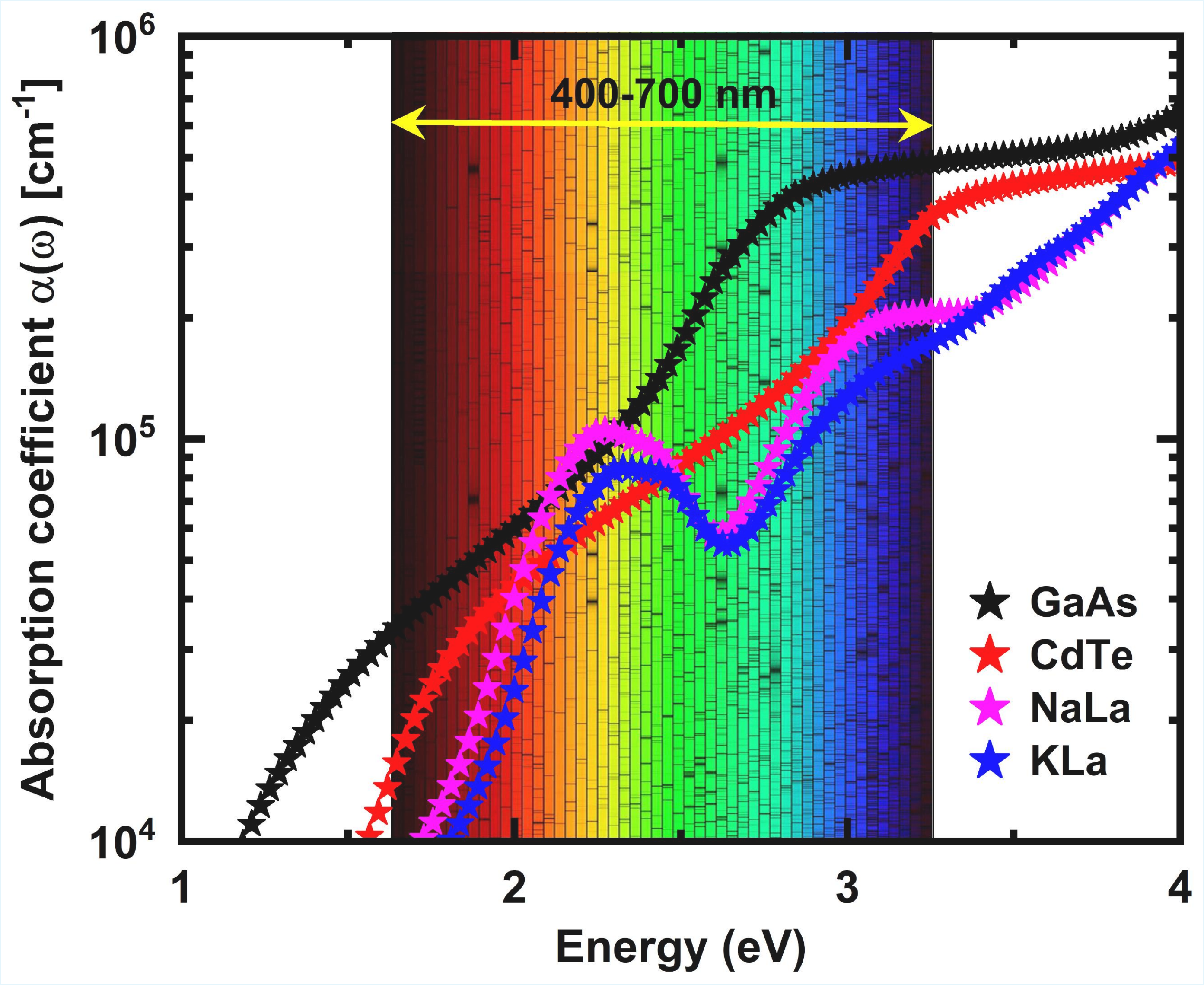}\vspace{-0pt}
\caption {Calculated optical absorption spectrum $\alpha(\omega)$ of AA$^\prime$FeMoO$_6$ (AA$^\prime$ = KLa and NaLa), compared with conventional semiconductors GaAs and CdTe.} 
\label{Figure3}
\end{figure} 
\par
To assess the optical response and light-harvesting potential, the absorption coefficient $\alpha(\omega)$ was calculated from the frequency-dependent complex dielectric function as depicted in \textcolor{blue}{Fig. \ref{Figure3}}. The $\alpha(\omega)$ spectra of KLaFeMoO$_6$ and NaLaFeMoO$_6$ show sharp optical onsets at 1.87 eV and 1.82 eV, respectively, positioning them in the optimal range for visible-light harvesting, superior to well-known BPVE material BiFeO$_3$ \cite{basu2008photoconductivity,wu2025enhancement}, and comparable with traditional PV semiconductors like GaAs and CdTe \cite{kato2020very,chen2018sustainable,shirayama2016optical}, as shown in \textcolor{blue}{Fig. \ref{Figure3}}. Above 2 eV, both materials exhibit $\alpha(\omega)$ value exceeding 1 × 10$^5$ cm$^{-1}$, ensuring strong light–matter interaction and efficient electron–hole generation. The redshift and increased $\alpha(\omega)$ observed in NaLaFeMoO$_6$ relative to KLaFeMoO$_6$ are attributed to its smaller band gap and high DOS of hybridized Fe-O in VB and Mo-O states in CB near the Fermi level (Refer SM \textcolor{blue}{Fig. S4b}). In both materials, the first sharp absorption peak arises from transitions between these strongly coupled states. This combination of strong visible-light absorption and spontaneous polarization facilitates high photocurrent generation together with efficient BPVE-driven carrier separation, highlighting their potential for FE–PV devices.
\begin{figure}
\centering
\includegraphics[width=\linewidth]{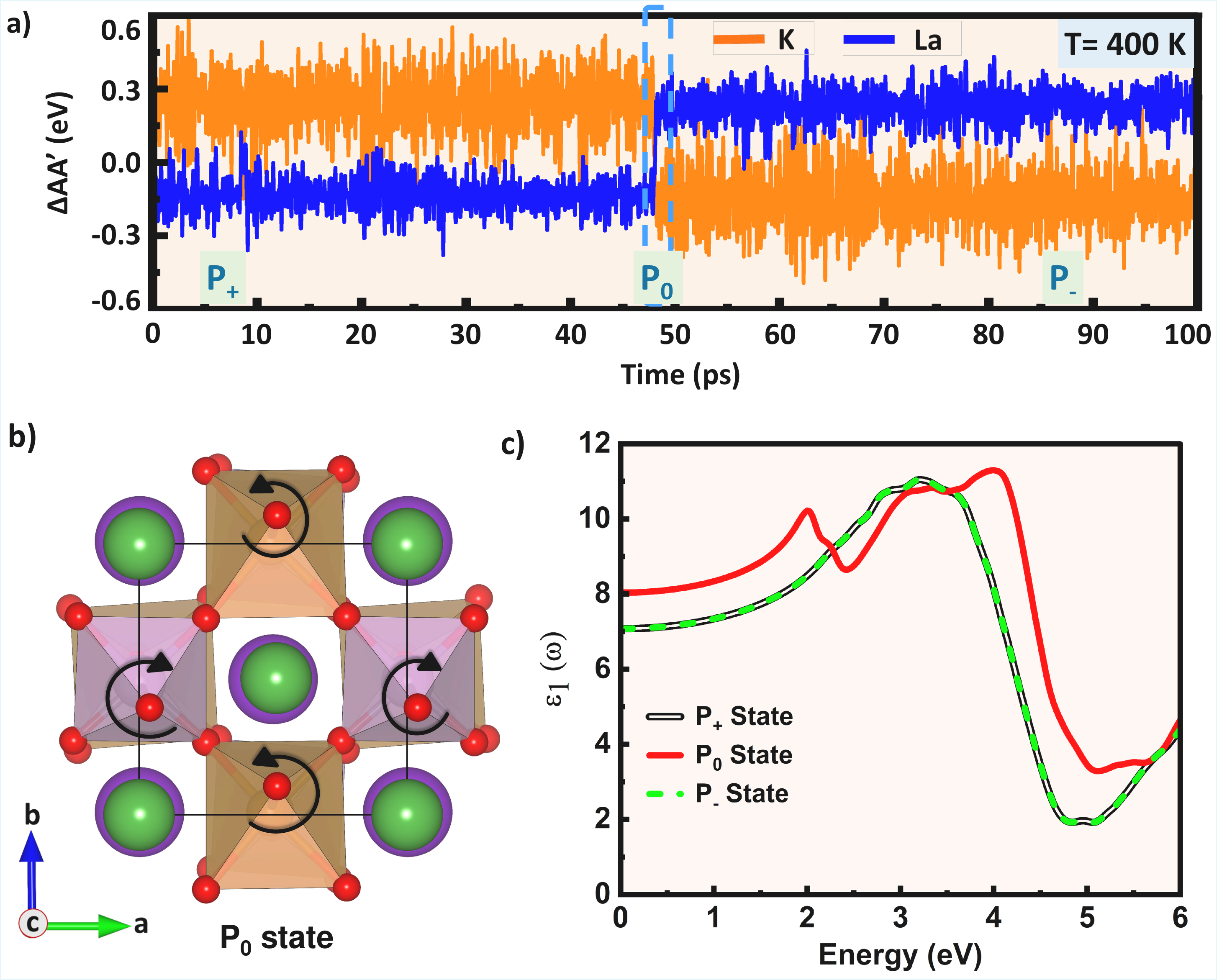}\vspace{-0pt}
\caption {(a) Displacement of K and La atoms over 100 $ps$ at 400 K. Snapshot of the structure in the intermediate polarization state ($\vec{P}_{0}$). (c) Real part of the dielectric function $\epsilon_{1}(\omega)$ for the polarization states (${P_{+}}$), (${P_{0}}$) and (${P_{-}}$).} 
\label{Figure4}
\end{figure}
\par
\textit{\textbf{Ferroelectric operating temperature (T$_S$)}}: The above studies have highlighted the potential of these materials as single-phase FE–PV candidates. However, for practical applications, stability of the FE phase at or above room temperature is essential to sustain polarization and ensure a steady photocurrent. If polarization becomes unstable or reverses spontaneously, persistent photocurrent cannot be maintained \cite{brody1993decay}. To identify the temperature range over which polarization remains stable, we assessed the ferroelectric operating temperature (T$_s$) using ab-initio molecular dynamics (AIMD) simulations. A \(2\times2\times2\) (160-atom) supercell of NaLaFeMoO$_6$ and KLaFeMoO$_6$ was used and the displacement of A (K and Na) and A$^\prime$ (La) cations from their centrosymmetric positions $\Delta$AA$^\prime$(\AA), as a function of time steps $\Delta$t ($ps$) at different temperatures was recorded during the simulations, are presented in \textcolor{blue}{Fig. S6 and S7} of the SM \cite{SM}. 
\par
For the KLaFeMoO$_6$ system, at 300 K, A(K) and A$^\prime$(La) site cations exhibit displacements, and the corresponding structure remains in the polar P2$_1$ phase with the $a^{-}a^{-}c^{+}$ tilt pattern and polarization $\Vec{P_{y}}$ aligned along [010] ($\Vec{P_{+}}$ state). When the temperature is increased to 400 K, the system initially retains the $\Vec{P_{+}}$ state. However, around 48 $ps$, the K and La cations acquire sufficient thermal energy to overcome the potential barrier and shift their positions as shown in the \textcolor{blue}{Fig. \ref{Figure4}a}. Structural analysis in the 48–100 $ps$ window shows that the system still belongs to the P2$_1$ phase but adopts a different tilt pattern $a^{-}a^{-}(-c)^{+}$, in which the entire in-phase tilt pattern reverses from $a^{-}a^{-}c^{+}$ to $a^{-}a^{-}(-c)^{+}$ along the $c$ axis. This transition confirms polarization switching in rotationally induced HIF materials \cite{mulder2013turning,feng2014high}. Specifically, the polarization switches from [010] ($\Vec{P_{+}}$ state) to [0$\Bar{1}$0] ($\Vec{P_{-}}$ state) at 400 K ($\sim$48 ps). \textcolor{blue}{Fig. S8} of the SM \cite{SM} shows snapshots of the $\vec{P}{_+}$, $\vec{P}{_0}$, and $\vec{P}{_-}$ states. The $\Vec{P_{0}}$ state preserves the tilt but loses the in-phase rotation, leading to negligible AA$^\prime$-cation displacements as shown in \textcolor{blue}{Fig. \ref{Figure4}b}. In the case of NaLaFeMoO$_6$, polarization switching occurs at 700 K, indicating a significantly higher T$_s$. This behavior arises from the smaller ionic radius mismatch between Na and La, which reduces A-site off-centering and thereby increases the energy barrier for polarization reversal, consistent with previous reports \cite{buvaneswaran2025ferroelectric,buvaneswaran2023design}.
\par
\textit{\textbf{Switching pathway}}: To gain insight into the microscopic switching mechanism, we analyzed the structural evolution of KLaFeMoO$_6$ during polarization reversal. The system initially adopts the in-phase tilt configuration ($a^{-}a^{-}c^{+}$), with anti-phase octahedral tilts along the $a$- and $b$-axes and an in-phase tilt along the $c$-axis. Simultaneously, the octahedra exhibit an in-phase rotation around the $c$-axis ($a^0a^0c^{+}$), which stabilizes the initial polarization and magnetization. During switching, the structure evolves toward the out-of-phase tilt configuration ($a^{-}a^{-}c^{0}$), retaining the anti-phase tilts along $a$ and $b$ but losing the $c$-axis tilt. Remarkably, the in-phase rotation ($a^0a^0c^{+}$) persists, maintaining a finite rotational component that stabilizes the system. This mode of switching, in which the tilt gradually changes from in-phase to out-of-phase while the in-phase rotation remains unchanged, is referred to as tilt precession Q$_{Tp}$ \cite{zanolli2013electric}. A snapshot of this pathway is provided in \textcolor{blue}{Fig. S8} of the SM \cite{SM}. The persistence of $Q_{R+}$ coupled with Q$_{Tp}$ plays a critical role in stabilizing both the polarization and magnetization vectors, ensuring an energetically favorable transition. As shown in \textcolor{blue}{Fig.\ref{Figure4}a}, around 48 $ps$, not only polarization switches, but also polarization-coupled magnetization switching is observed. A noncollinear configuration characterized by F$_x$A$_y$G$_z$ shows reversal of the weak ferromagnetic component along $x$ ($+F_x \xrightarrow{}-F_x$, with magnetization along $x$, $m_x$ = 0.35 $\mu_B$/Fe driven by polarization reversal ($+P_y \xrightarrow{}-P_y$, with polarization along $P_y$ = 21.32 $\mu$C/cm$^2$).
\par
The dielectric response, shown in \textcolor{blue}{Fig.\ref{Figure4}c}, provides an additional probe to track polarization switching. At 400 K, both the P$_{+}$ (0–48 ps) and P$_{-}$ (48–100 ps) states exhibit nearly identical $\epsilon_{1}(\omega)$ spectra, with a static dielectric constant of 7.07, indicating that the reversal of octahedral tilt orientation alone does not significantly affect the optical response. In contrast, the transient P$_{0}$ state ($\sim$48 ps) shows a clear enhancement, with the static dielectric constant increasing to 8.03. This enhancement results from the suppression of in-phase rotation, which eliminates ionic polarization but strengthens Fe–O/Mo–O orbital overlap. The resulting band-gap narrowing increases the joint DOS near the band edges, thereby amplifying optical transitions. Thus, the dielectric spectrum not only reflects electronic structure changes but also serves as a reliable indicator of polarization switching, revealing the strong coupling between polarization dynamics, band structure, and dielectric response.
\par
In conclusion, we proposed and characterized A/A$^\prime$-ordered double perovskites KLaFeMoO$_6$ and NaLaFeMoO$_6$ as promising single-phase ferroelectric photovoltaics. They exhibit A-site [L] and B-site [R] ordering with P2$_1$ symmetry, $a^{-}a^{-}c^{+}$ tilts, and significant polarization. Both compounds have indirect band gaps (1.87 and 1.82 eV), strong visible-light absorption ($>$10$^5$ cm$^{-1}$), and low effective masses along the polar axis, favorable for efficient charge transport. AIMD simulations confirm that both materials maintain a stable ferroelectric phase at or above room temperature, with tilt-precession as the primary pathway for polarization and magnetization reversal. These results highlight their potential for next-generation ferroelectric photovoltaic and multifunctional optoelectronic applications.
\begin{acknowledgments}
S.G. acknowledges funding from DST-ANRF Core Research Grant File. no. CRG/2023/3209 for funding. 
S.B. acknowledges the SRMIST KTR for his fellowship. 
\end{acknowledgments}

%
\end{document}